\documentclass{elsart}

\usepackage{amssymb}
\usepackage{graphicx}

\begin{document}

\begin{frontmatter}



\title{Shell gaps and $pn$ pairing interaction in $N=Z$ nuclei}


\author[a1]{Kazunari~Kaneko}
\author[a2]{Jing-Ye~Zhang}
\author[a3]{Yang Sun}

\address[a1]{Department of Physics, Kyushu Sangyo University,
Fukuoka 813-8503, Japan}
\address[a2]{Department of Physics and Astronomy, University
of Tennessee, Knoxvill, Tennessee 37996, USA}
\address[a3]{Department of Physics, Shanghai
Jiao Tong University, Shanghai 200240, P. R. China}

\begin{abstract}
We analyze the observed shell gaps in $N=Z$ nuclei determined from
the binding energy differences. It is found that the shell gaps can
be described by the combined contributions from the single-particle
level spacing, the like-nucleon pairing, and the proton-neutron
pairing interaction. This conclusion is consistent with that of
Chasman in Phys. Rev. Lett. {\bf 99} (2007) 082501. For the
double-closed shell $N=Z$ nuclei, the single-particle level spacings
calculated with Woods-Saxon potential are very close to those
obtained by subtracting the $nn$ pairing interaction from the
observed shell gap. For the sub-closed or non-closed shell $N=Z$
nuclei, the $pn$ pairing interaction is shown to be important for
the observed shell gaps.
\end{abstract}

\end{frontmatter}


A long-standing problem in nuclear structure is that for the
double-closed and sub-closed shell nuclei the single-particle
energy-level orderings and spacings obtained from mean-field
calculations underestimate the observed shell gaps. The observed
shell gaps are defined by taking differences of ground-state masses,
which are usually given as twice the odd-even mass difference
extracted from the binding energy. However, this method assumes that
there are no many-body effects involved in the mass differences at
the closed shell. Chasman \cite{Chasman} has recently investigated
this problem, and pointed out that the correlation energy due to
pairings can resolve this discrepancy. As addressed in
Ref.\cite{Chasman}, there are changes in binding energy due to
many-body effects even for double-closed shell nuclei. In this
papper, two main interactions have been considered to affect the
observed shell gaps: One is the pairing interaction in like nucleons
(neutron-neutron($nn$) and proton-proton ($pp$)), and the other is
the proton-neutron ({\it pn}) pairing interaction. Previously, one
of us (K.\ K.) studied \cite{kaneko99,kaneko01,kaneko04} these
empirical interactions in $N\approx Z$ nuclei using the odd-even
mass difference and the double difference of binding energies
\cite{Janecke3,Jensen}, which is different from the other one
\cite{Jingye}.

A typical indicator for $T=1$ $nn$ pairing interactions is well
known, which is given by the following three-point odd-even mass
difference:
\begin{eqnarray}
\Delta^{(3)}_{n}(Z,N) & = & \frac{(-1)^{N}}{2}[B(Z,N+1) \nonumber \\
                      & - & 2B(Z,N)+B(Z,N-1)],
\label{eq:1}
\end{eqnarray}
where $B(Z,N)$ is the negative binding energy of a nucleus.
According to the standard BCS theory for the $nn$ pairing gap
$\Delta_{n}$, $B(Z,N\pm 1)\approx B(Z,N)+\Delta_n \pm \lambda_n$.
Therefore, $\Delta^{(3)}_{n}(Z,N)$ is roughly $\Delta_{n}$. Thus,
$\Delta^{(3)}_{n}$ is often interpreted as a measure of the
empirical $nn$ pairing gap. However, because of the odd-even
staggering effect, values of $\Delta^{(3)}_{n}(Z={\rm even},N)$ are
large for even-$N$ and small for odd-$N$ nuclei. It has been
suggested \cite{Satula2,Dobaczewski} that the three-point odd-even
mass difference for an odd-mass nucleus with neutron excess is an
excellent measure of $pp$ and $nn$ pairing interactions in
neighboring even-even nucleus, although it is still controversial
\cite{Duguet}. Thus, the differences of $\Delta^{(3)}_{n}$ in
adjacent even- and odd-$N$ nuclei reflect the mean-field
contributions. To extract the mean-field shell gap, we can apply
this idea to the even-even $N=Z$ nuclei.

\begin{figure}[t!]
\includegraphics[totalheight=5.4cm]{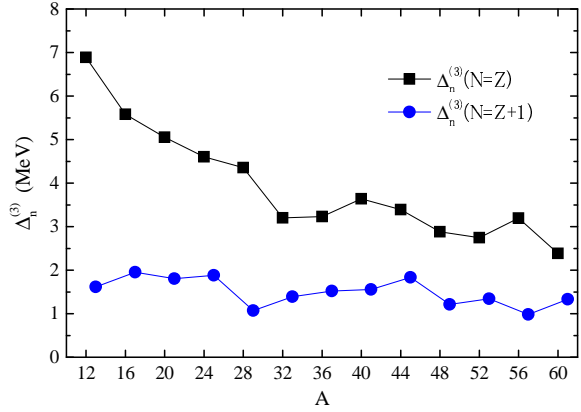}
\caption{(Color online) Experimental odd-even mass differences
$\Delta^{(3)}_{n}(Z,Z)$ for even-even $N=Z$ nuclei, and
$\Delta^{(3)}_{n}(Z,Z+1)$ for the neighboring odd-mass $N=Z+1$
nuclei, plotted as functions of mass $A=2Z$ and $A=2Z+1$,
respectively.} \label{fig1}
\end{figure}

Figure \ref{fig1} shows experimental values of $\Delta^{(3)}_{n}$
obtained by using Eq. (\ref{eq:1}). We plot $\Delta^{(3)}_{n}(Z,Z)$
and $\Delta^{(3)}_{n}(Z,Z+1)$ for the even-even $N=Z$ and the
adjacent odd-mass $N=Z+1$ nuclei ranging from $A=$12 to $A=$61. It
can be seen that the large $\Delta^{(3)}_{n}(Z,Z)$ in $N=Z$ nuclei
decreases steadily with increasing particle number. The expected
quenching in the $nn$ pairing interaction at the magic or semi-magic
number $N$ or $Z$ = 14 and 28 is clearly seen. As also observed from
the figure, the differences between $\Delta^{(3)}_{n}(Z,Z)$ and
$\Delta^{(3)}_{n}(Z,Z+1)$ are remarkable. The $N=Z$ nuclei have
additional binding energy due to the so-called Wigner effect. The
differences in $\Delta^{(3)}_{n}$ between neighboring even- and
odd-$N$ nuclei reflect the single-particle (mean-field)
contributions and the correlation energies.

To investigate the physical source of the differences between
$\Delta^{(3)}_{n}(Z,Z)$ and $\Delta^{(3)}_{n}(Z,Z+1)$, we first
adopt a spherical single-particle model without considering the
two-body interactions. In this case, the binding energy is simply
expressed as
\begin{eqnarray}
B_{sp}(N) & = & \sum_{j}N_{j}\epsilon_{j},
\label{eq:2}
\end{eqnarray}
with $N_{j}$ and $\epsilon_{j}$ being the occupation number and
single-particle energy, respectively, and the particle number
$N=\sum_{j}N_{j}$. In a double-closed and sub-closed shell nucleus
with $N=Z$, energy levels are fully occupied up to the level
$j_{0}$, while in the neighboring odd-$N=Z+1$ system, the last
neutron occupies the next level $j_{1}$. This implies
\begin{eqnarray}
\Delta^{(3)}_{sp}(N=Z+1) & = & 0, \nonumber \\
\Delta^{(3)}_{sp}(N=Z) & = & \frac{1}{2}(\epsilon_{1}-\epsilon_{0}),
\label{eq:3}
\end{eqnarray}
where $\epsilon_{0}$ and $\epsilon_{1}$ are the level energies for
$j_0$ and $j_1$, respectively. Thus, for the double-closed and
sub-closed shell nuclei with $N=Z$, the indicator (\ref{eq:1})
vanishes for $N=Z+1$, but gives half of the single-particle level
spacings for $N=Z$. On the other hand, for non-closed shell nuclei
with $N=Z$ particles partially occupy the last level $j_{0}$. The
odd-even mass difference is then expressed as
\begin{eqnarray}
\Delta^{(3)}_{sp}(N=Z+1) & = & 0, \nonumber \\
\Delta^{(3)}_{sp}(N=Z) & = & 0.
\label{eq:4}
\end{eqnarray}
This means that the single-particle energies do not contribute to
the odd-even mass difference or the observed shell gap for
non-closed shell nuclei with $N=Z$. It is important to note that
polarization effects for odd-$A$ nucleus may affect the filters
(\ref{eq:3}) and (\ref{eq:4}). For this reason, the formulae
(\ref{eq:3}) and (\ref{eq:4}) are considered as approximations.

The many-body contributions beyond the single-particle model are
characterized by the amount that deviates from Eq. (\ref{eq:3}). By
subtracting the many-body contributions from the indicator
(\ref{eq:1}) at $N=Z$ and $N=Z+1$, we may obtain information about
the single-particle level spacing. Since for non double-closed or
non sub-closed shell nuclei both values in (\ref{eq:3}) vanish in a
single-particle model, the many-body contributions are dominated by
the odd-even mass difference in these nuclei.

\begin{figure}[t!]
\includegraphics[totalheight=5.0cm]{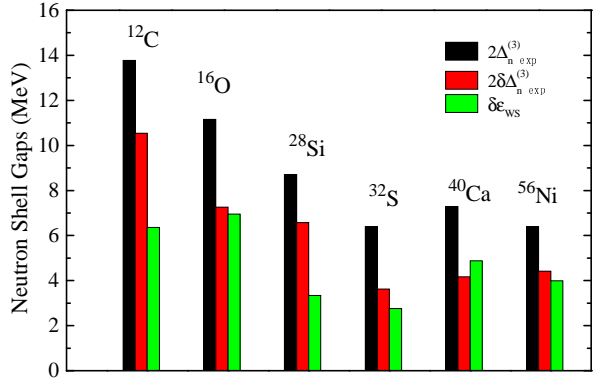}
\caption{(Color online) Comparison of shell gaps in magic and
submagic nuclei. For each nucleus, the quantities are ordered (from
left to right) as the observed gap $2\Delta^{(3)}_{n}(Z,Z)$, the
extracted gap $2\delta \Delta^{(3)}_{n}(Z,Z)$, and the calculated WS
spacing $\delta \varepsilon_{WS}$.} \label{fig2}
\end{figure}

We consider the observed shell gap defined as twice the odd-even
mass difference $\Delta^{(3)}_{n}(Z,Z)$. By subtracting the $nn$
pairing gap $\Delta^{(3)}_{n}(Z,Z+1)\approx \Delta_{n}$ from
$\Delta^{(3)}_{n}(Z,Z)$, we can define the extracted gap as
\begin{eqnarray}
\delta \Delta^{(3)}_{n}(Z,Z) & = & \Delta^{(3)}_{n}(Z,Z) -
\Delta^{(3)}_{n}(Z,Z+1).
\label{eq:5}
\end{eqnarray}
In Fig. \ref{fig2}, we plot twice the observed shell gap
$2\Delta^{(3)}_{n}(Z,Z)$, twice the extracted gap $\delta
\Delta^{(3)}_{n}(Z,Z)$, and the single-particle spacing $\delta
\varepsilon_{WS}$, for the double-closed and sub-closed shell nuclei
ranging from $A=$12 to $A=$56. Comparing the extracted gaps with the
single-particle spacings $\delta \varepsilon_{WS}$ obtained from a
Woods-Saxon (WS) potential, we can see that the agreement between
these two quantities is fairly good for the double-closed shell
nuclei $^{16}$O, $^{40}$Ca, and $^{56}$Ni. This is expected because
the $nn$ pairing interaction is dominated in these nuclei and any
other interactions would be small due to the large shell gaps. Thus,
for the double-closed shell nuclei, the $nn$ pairing interaction is
considered to be the extracted gaps $\delta\Delta^{(3)}_{n}(Z,Z)$.
For sub-closed shell nuclei such as $^{12}$C and $^{28}$Si, however,
the difference between the extracted gaps and the Woods-Saxon
calculations, defined as
\begin{eqnarray}
\delta \Delta_{n}(Z,Z) & = & \delta\Delta^{(3)}_{n}(Z,Z) -
\frac{1}{2}\delta \varepsilon_{WS},
\label{eq:6}
\end{eqnarray}
is quite large. This suggests that the many-body interactions beyond
the $nn$ pairing interaction would be significant. It should be
mentioned here that the WS potential model is by no means a
consistent microscopic mean-field model. A recent paper
\cite{Zalewski} has demonstrated that the single-particle energies
can be improved systematically by refitting the spin-orbit and
tensor part of the energy density functional method. Inclusion of
the tensor effect may modify the shell gaps in sub-closed-shell
nuclei.

\begin{figure}[b!]
\includegraphics[totalheight=6.4cm]{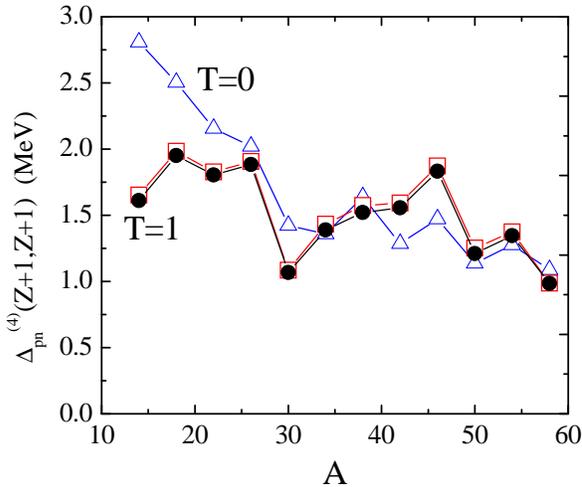}
\caption{(Color online) The $pn$ pairing gaps estimated from the
double differences of experimental binding energies. The open
triangles denote the $T=0$ $pn$ pairing gap, while the solid circles
the $T=1$ $pn$ pairing gap. The odd-even mass differences in
odd-mass nuclei with $N=Z+1$ are shown by the open squares. }
\label{fig3}
\end{figure}

Next, we study the many-body effects in the shell gaps for the
sub-closed and non-closed shell $N=Z$ nuclei. The odd-even mass
differences in even-even $N=Z$ nuclei are larger than those in the
neighboring even-even $N=Z+2$ nuclei, which reflects the gain in
pairing energy due to stronger $pn$ interactions in $N=Z$ systems
\cite{kaneko99} and is referred as the Wigner energy \cite{Satula1}.
The previous work \cite{kaneko99} studied the indicator (\ref{eq:1})
for the Cr isotopes by performing shell model calculations, and
suggested that the $pn$ pairing interactions play an important rule
for the odd-even mass difference as well as the $nn$ pairing at
$N=Z$. To describe the $pn$ pairing interactions in odd-odd $N=Z$
nuclei, we estimate the following double difference of binding
energies \cite{Janecke3,Jensen,kaneko99,kaneko01,kaneko04}:
\begin{eqnarray}
\Delta_{pn}^{(4)T}(Z,N) & = & \frac{(-1)^{N}}{2}[B(Z,N)^{T} - B(Z,N-1)
\nonumber \\
   & - &  B(Z-1,N)+B(Z-1,N-1)], \label{eq:7}
\end{eqnarray}
where $B(Z,N)^{T}$ is the binding energy of the lowest state of
isospin $T$ in odd-odd $N=Z$ nuclei. Figure \ref{fig3} shows such
double differences calculated from the experimental binding
energies. The odd-even mass differences for odd-mass nuclei are also
displayed. One sees that $\Delta^{(3)}_{n}(Z={\rm even},Z+1)$ agrees
with $\Delta_{pn}^{(4)T=1}(Z+1,Z+1)$, which means that the $T=1$
$pn$ pairing interaction for odd-odd $N=Z$ nuclei have the same
interaction energy as the $nn$ pairing interaction, namely
$\Delta_{n}=\Delta_{pn}^{(4)T=1}$, if isospin symmetry is assumed.
Thus, the indicator $\Delta_{pn}^{(4)T=1}$ provides the $T=1$ $pn$
pairing gap in $N=Z$ nuclei. Similarly, $\Delta_{pn}^{(4)T=0}$ can
be regarded as the $T=0$ $pn$ pairing gap. Figure \ref{fig3} further
suggests that for the ground states of $sd$ shell nuclei, the $T=0$
$pn$ interactions are stronger than the $T=1$ $pn$ interactions,
whereas an opposite situation occurs in the $pf$ shell nuclei where
the $T=1$ $pn$ interactions are stronger. Thus, the $T=0$ $pn$
pairing gap $\Delta_{pn}^{(4)T=0}$ cannot be explained by the $T=1$
pairing Hamiltonian.

\begin{figure}[t!]
\includegraphics[totalheight=7.4cm]{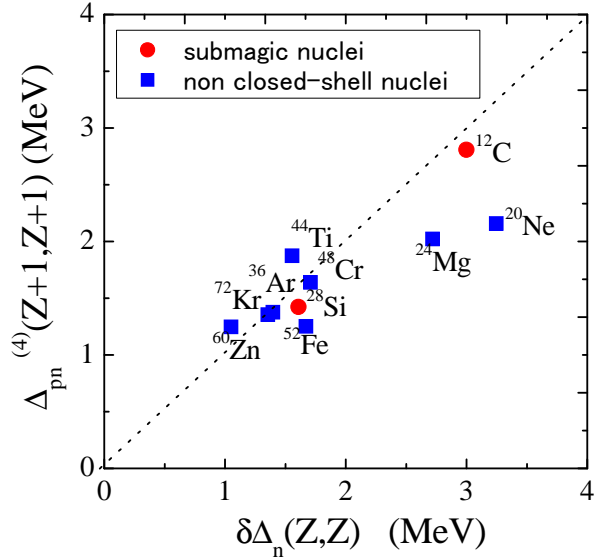}
\caption{(Color online) Comparison of the extracted gap with the
$pn$ pairing interaction. The solid squares and solid circles are for non-closed
and sub-closed shell nuclei, respectively. For the extracted gap of
sub-closed shell nuclei, we subtracted the WS part from the
extracted gap $\delta \Delta_{n}(Z,Z)$, using the equation
(\ref{eq:6}).} \label{fig4}
\end{figure}

In Fig. \ref{fig4}, we compare the extracted gap obtained from Eq.
(\ref{eq:5}) with the $pn$ pairing gap $\Delta_{pn}^{(4)T}(Z+1,Z+1)$
after subtracting the WS single-particle spacing from $\delta
\Delta^{(3)}_{n}(Z,Z)$. One can see that overall, the extracted gaps
correlate fairly well with the $pn$ pairing interaction, with only
two exceptions $^{20}$Ne and $^{24}$Mg (see discussions below).
Thus, we may conclude that $\Delta^{(3)}_{n}(Z,Z)$ generally
contains contributions from the single-particle spacing $\delta
\varepsilon$, the $nn$ pairing gap $\Delta_{n}$, and the $pn$
pairing gap $\Delta_{pn}$:
\begin{eqnarray}
\Delta^{(3)}_{n}(Z,Z) & = & \frac{1}{2}\delta \varepsilon +
\Delta_{n} + \Delta_{pn}. \label{eq:8}
\end{eqnarray}
It is important to note that $\Delta_{pn}\approx 0$ for the
double-closed shell nuclei and $\delta \varepsilon\approx 0$ for the
non-closed shell nuclei.

\begin{figure}[b!]
\includegraphics[totalheight=5.0cm]{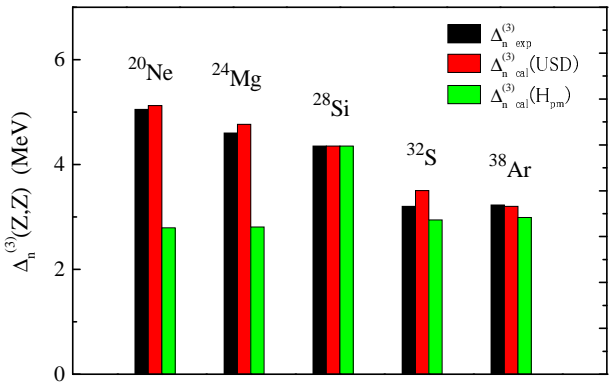}
\caption{(Color online) Comparison of the odd-even mass difference
$\Delta^{(3)}_{n}(Z,Z)$ for $N=Z$ nuclei in the sd-shell. For each
nuclide, the ordering is the experimental values, the calculated
values in the USD interaction, and the calculated values in the
$H_{pm}$ interaction.} \label{fig5}
\end{figure}

Using shell model calculations with the USD interaction in the sd
shell, we now explain why the extracted gaps of $^{20}$Ne and
$^{24}$Mg in Fig. \ref{fig4} are larger than the $pn$ pairing
interaction. To understand this, we first extract the $J=0$ pairing
and monopole interaction from the USD interaction \cite{USD}
\begin{eqnarray}
H_{pm} & = & H_{0} + H_{p} + H_{m}. \label{eq:9}
\end{eqnarray}
In Eq. (\ref{eq:9}), $H_{0}$ is the single-particle Hamiltonian and
the pairing term $H_{p}$ has the $J=0$ components of the two-body
matrix elements $\langle a,b,J,T|V|a,b,J,T \rangle$ in the USD
interaction. Hence the matrix elements of the monopole interaction
$H_{m}$ take the form
\begin{eqnarray}
V_{m}^{T}(a,b) & = & \frac{\sum_{J}(2J+1)\langle a,b,J,T|V|a,b,J,T
\rangle}{\sum_{J}(2J+1)},
\label{eq:10}
\end{eqnarray}
where $a, b$ are single particle orbitals and the $J=0$ components
are neglected from the summation. The residual interaction is then
defined by $H_{res}=H - H_{pm}$. It is well known that this
interaction is dominated by the multipole interactions such as the
quadrupole, octupole, and hexadecapole interactions \cite{Dufour}.
In this sense, the residual interaction $H_{res}$ provides the
collective correlations \cite{hasegawa,kaneko02}. On the other hand,
the monopole interaction does not lead to the collective
correlations but it is important for the binding energy. It has been
shown \cite{kaneko99,kaneko01,kaneko04} that the $T=0$ matrix
elements of the monopole field $V_{m}^{T}(a,b)$ are significantly
larger than those with $T=1$, and are very important in determining
the double differences of binding energies
\cite{Jensen,Jingye,Satula2}. We can see that the matrix elements
are quite large for the isoscalar components but small for the
isovector components \cite{kaneko01}. In the USD interaction, the
monopole matrix elements (\ref{eq:10}) with $T=0$ have values about
$-3$ MeV and are strongly attractive, while the $T=1$ monopole
components are quite small.

The experimental and theoretical odd-even mass differences
$\Delta^{(3)}_{n}(Z,Z)$ for $N=Z$ nuclei in the sd-shell region are
compared in Fig. \ref{fig5}. The calculated results with the USD
interaction reproduce very well the odd-even mass difference for the
$N=Z$ nuclei. The shell model calculations with the $J=0$ pairing
and monopole interactions are in good agreement with the
experimental values for $^{28}$Si, $^{32}$S, and $^{36}$Ar, but not
for $^{20}$Ne and $^{24}$Mg. It is obvious that the differences for
$^{20}$Ne and $^{24}$Mg are attributed to the residual interaction
$H_{res}$, and are consistent with the previous finding that the
extracted gaps are larger than the $pn$ pairing interactions in Fig.
\ref{fig4}.

To summarize, we have studied in detail the observed shell gaps
determined from the binding energy differences for $N=Z$ nuclei. We
have shown that the observed shell gaps can be described by the
single-particle level spacing, the $nn$ pairing interaction, and the
$pn$ pairing. This conclusion is consistent with that of Chasman
\cite{Chasman}. In particular, the $pn$ pairing interactions are
important for the non-closed and sub-closed shell nuclei, while they
can be neglected for double-closed shell nuclei. For $^{20}$Ne and
$^{24}$Mg, it has been found that the residual interactions after
removing the $J=0$ pairing and monopole interactions contribute to
the shell gaps as well. Although we have considered in this Letter
the neutron shell gap only, similar conclusions for the proton shell
gap can also be obtained.

Y.S. was supported in part by the Chinese Major State Basic Research
Development Program through grant 2007CB815005.



\end{document}